\documentclass[aps,prl,preprint,showpacs,preprintnumbers,amsmath,amssymb]{revtex4}
\usepackage{graphicx}
\usepackage{txfonts}

\begin{document}


\title{Proposed measurements of the interlayer magnetoresistance of
   underdoped cuprate superconductors can distinguish closed pockets from
   open arcs in the Fermi surface}

 \author{M. F. Smith}
 \email{mfsmith@g.sut.ac.th}
\affiliation{
School of Physics, Suranaree University of Technology, Nakhon Ratchasima 30000, Thailand }
\affiliation{
Department of Physics, University of Queensland,  Brisbane 4072, Australia}

\author{Ross H. McKenzie}
\affiliation {Department of Physics, University of Queensland,
 Brisbane 4072,  Australia}

\date{\today}

\begin{abstract}
An outstanding question concerning the underdoped cuprate concerns the true
nature of their  Fermi surface which appears as a set of disconnected arcs.
Theoretical models have proposed two distinct possibilities: (1)
each arc is the observable part of a partially-hidden closed pocket,
and (2) each arc is open, truncated at its apparent ends.
We show that measurements of the variation of
the interlayer resistance with the direction of a magnetic field parallel to
 the layers can qualitatively distinguish closed pockets from open arcs.
  This is possible because the field can be oriented such that all electrons on arcs encounter a large Lorentz force and resulting magnetoresistance whereas some electrons on pockets escape the effect by moving parallel to the field.

\end{abstract}

\maketitle

 The Fermi surface (FS) of underdoped cuprates in the pseudogap state appears, in electronic spectrum measurements, as four short arcs near diagonals of the Brillouin zone \cite{kond09,lee07,dama03,mars96,norm05,jlee09,kohs08,push09}.  These arcs neither close back on themselves nor terminate at zone boundaries, which are the only possibilities for a conventional FS, but rather
 end abruptly within the zone interior.  
 According to some theoretical pictures\cite{hur09,xia09,yang06,yang09,ng05}, each apparently open spectral arc is just the observable segment of a closed Fermi surface pocket (the missing side of the pocket is claimed to be present but undetected because of its lower spectral weight).  In contrast, others
 propose that truly open arcs, without any closed pockets, comprise the FS.\cite{lee06,lee09,norm07}
In this Letter, we show that the interlayer magnetoresistance (IMR)
is {\it qualitatively different} for closed pockets and open arcs. Hence, the IMR measurements we propose should be able to rule out a whole class of theoretical models for the pseudogap state.

 Though quasiparticle peaks on the arcs are broad in zero magnetic field, the observation of quantum oscillations (QOs) in underdoped cuprates\cite{doir07, lebo07,bang08,jaud09,seba08} indicates that sharp quasiparticles are present in
high fields.  Based on their frequency, the oscillations may be plausibly attributed to quasiparticles on the spectral arcs\cite{AFMnote} but either closed pockets or open arcs\cite{pere09} can accommodate QOs.  To elucidate the connection between QOs and the nature of the spectral arcs
we need a complementary probe, one that accesses the high-field phase where QOs are seen and determines whether the quasiparticles more likely live on a closed or open FS.

The dependence of the IMR on the
direction of the magnetic field has proven to be a powerful probe
of Fermi surface properties in overdoped cuprates \cite{huss03,AbdelNP06,Kennett}.  We have previously proposed that it can be used to map the anisotropy of a weak pseudogap\cite{smit09}.   
Significant IMR effects require a magnetic field strong enough that the cyclotron frequency $\omega_C$ is of order the scattering rate $\tau^{-1}$,
the same condition
needed for QOs\cite{omtau}.  When the field ${\bf B}$ is in the conducting layers, only quasiparticles moving parallel to
${\bf B}$, which feel no Lorentz force, avoid a large classical magnetoresistance to interlayer current.  Two classes of FS can be distinguished by their qualitatively different ${\bf B}$ dependencies.  In the first, a quasi-2D system, there are certain to be quasiparticles somewhere on the FS with velocity parallel to any particular ${\bf B}$.  In the second, that of quasi-1D metals, it is possible to choose a ${\bf B}$ along which no quasiparticles are moving.    We argue that Fermi pockets fall into the first (2D) class of FS and open Fermi arcs into the second (1D) class, so that they may be distinguished by IMR.  We discuss potential complications below after describing the effect in more detail.

 A magnetic field
 ${\bf B}=B_0(\cos\phi_B,\sin\phi_B,0)$
applied within the conducting layers can be described by a vector potential
 ${\bf A}= z {\hat z} \times {\bf B}$
that depends on interlayer position $z$.
The IMR $\rho_{zz}(B)$ is:
\begin{equation}
\label{rho1}
\rho_{zz}^{-1}(B)=\frac{\mathrm{e}^2c}{\pi}\sum_\sigma\int d^2{\bf k} \ 
t_\perp^2({\bf k})\int d\omega\bigg{(}-\frac{df_0}{d\omega}\bigg{)}
\Pi_{12}({\bf k},\omega)
\end{equation}
where $f_0(\omega)$ is the Fermi function, and $\Pi_{12}({\bf k},\omega)=D_{1\sigma}({\bf k},\omega)D_{2\sigma}({\bf k},\omega)$ is the product of spectral functions on adjacent layers:
 $D_{1\sigma}({\bf k},\omega)$ is the spin-$\sigma$ spectral function for the $z=0$ layer and $D_{2\sigma}({\bf k},\omega)=
 D_{1\sigma}({\bf k} - e {\bf A},\omega)$
 the same for $z=c$ where $c$ is the interlayer spacing\cite{mahan,moses}.  The small interlayer hopping element $t_\perp({\bf k})$ depends strongly on ${\bf k}$ in the layer, we use\cite{ande95} $t_\perp({\bf k})=t_\perp(\cos k_x - \cos k_y)^2$ and work to lowest order in $t_\perp$.

Using a metallic spectral function with quasiparticle energies $E_{1\sigma{\bf k}}$ and $E_{2\sigma{\bf k}}$, on the two layers (both are shifted by the Zeeman energy $\mu_BB$) and scattering rate $\tau^{-1}$, Eq. (\ref{rho1}) becomes:
\begin{equation}
\label{rhoclean}
\rho_{zz}^{-1}(B)=\rho_{zz}^{-1}(0)\bigg{<}t_\perp^2({\bf k})[1+(\Lambda_{12}\tau)^2]^{-1}\bigg{>}_{FS}\bigg{<}t_\perp^2(\bf k)\bigg{>}_{FS}^{-1}
\end{equation}
where $\Lambda_{12}=E_{1{\bf k}}-E_{2{\bf k}}$ (the Zeeman terms cancel, so we drop the spin index), and angle brackets denote an average over the $E_{1{\bf k}}=0$ surface, i.e. $<f({\bf k})>_{FS}=\int d{\bf k}f({\bf k})\delta(E_{1{\bf k}})$.
  We have $\Lambda_{12}=\Omega_C\tau({\hat{\bf v}_{el}}\cdot{\hat{\bf A}})$ where ${\bf v}_{el}$ is the electric current velocity of the quasiparticle (proportional to its intralayer electric current) and $\Omega_C\tau=eB_0v_{el}c\tau$.  Equation (\ref{rhoclean}) is similar
to equations for normal metals\cite{kart04, scho97, huss96}; in this Letter we present a version relevant to Fermi arcs and pockets.

On a closed 2D FS, for any ${\bf B}$ there must be a set of FS points ${\bf k}^*$ at which ${\bf v}_{el} \parallel {\bf B}$.  For large fields, i.e. $\Omega_C\tau
\gg 1$, we expand around these FS points to find:
\begin{equation}
\label{rholin}
\rho_{zz}^{-1}(B)=\rho_{zz}^{-1}(0)\frac{\sum_{{\bf k}^*} F_{{\bf k}^*}(\Omega_C\tau)^{-1}\eta_{{\bf k}^*}}{<F_{\bf k}>_{FS}}.
\end{equation}
where $F_{{\bf k}}=t_\perp({\bf k})^2/|v_f|$ and $\eta_{{\bf k}^*}=1/2\Lambda_{12}^{''}$ with the second derivative of $\Lambda_{12}$ evaluated at ${\bf k}^*$ with respect to a vector perpendicular to the energy gradient.  The resistance is linear in field\cite{scho97,moses} for any orientation of ${\bf B}$.

If the FS is open (like in a quasi-1D metal) there are ${\bf B}$ for which no points on the surface satisfy ${\bf v}_{el}\parallel {\bf B}$.  For such ${\bf B}$, and $\Omega_C\tau>>1$, $\Lambda_{12}\tau$ is always large compared to unity so $\rho(B)\propto B^2$.  There are other directions of ${\bf B}$ for which ${\bf v}_{el}$ is nearly parallel to ${\bf B}$ over a wide slab of the FS, so that $\Lambda_{12}\tau$ is small and $\rho_{zz}(B)$ weakly $B$-dependent.

 To make the connection with the underdoped cuprates we consider the following model spectral function\cite{norm07,smit10} that captures pocket or arc models with appropriate parameter choices:
\begin{equation}
\label{specfun}
D_1({\bf k},\omega)=u_{\bf k}^2\frac{\gamma}{(\omega-E_{{\bf k}+})^2+\gamma^2}+v_{\bf k}^2\frac{\gamma}{(\omega-E_{{\bf k}-})^2+\gamma^2}
\end{equation}
where $E_{{\bf k}\pm}=\mu_{\bf k}\pm E_{\bf k}$, $E_{\bf k}=\sqrt{\xi_{\bf k}^2+\Delta_{\bf k}^2}$, $v_{\bf k}^2=(1/2)[1-\xi_{\bf k}/E_{\bf k}]$, $u_{\bf k}^2=1-v_{\bf k}^2$, $\gamma=1/(2\tau)$, $\xi_{\bf k}$ is a (normal metallic) band energy and $\Delta_{\bf k}$ is the pseudogap.

Closed Fermi pockets can be realized by taking $\mu_{\bf k}$ to be positive near nodes ${\bf k}_n$, which are located where $\xi_{\bf k}=0$ on the zone diagonal\cite{yang06,yang09,ng05,norm07}.  This gives a pocket Fermi surface $E_{{\bf k}-}=0$.  The spectral weight $v_{\bf k}^2$ suppresses one side of the pocket, making the model consistent with observed spectral arcs.  Assuming well-defined quasiparticles exist, $\gamma$ is smaller than relevant band parameters including $\mu_{{\bf k}_n}$.  The current is thus dominated by the band with pockets (the second term in Eq. (\ref{specfun})).

The {\it crucial property of pockets is that the current velocity} ${\bf v}_{el}={\bf \nabla}E_{{\bf k}-}$
{\it is normal to the pocket surface.} Every direction in the layer is represented by the velocity ${\bf v}_{el}$  somewhere on the pocket (see Fig. 1).  This is true despite the anisotropic spectral weight.  For, upon adding the total interlayer current of two pockets on opposite sides of the Brillouin zone, the spectral factors combine to give one full pocket out of the two partially hidden ones.  Any model with quasiparticle current that sweeps through all directions belongs to the quasi 2D class of FS to which Eq. (\ref{rholin}) applies.

Open Fermi arcs can be modeled using Eq. (\ref{specfun}) with the pseudogap taken to be zero in a range of directions near the diagonal, turning on suddenly at arc ends\cite{pere09}.  On arcs, $\xi_{\bf k}=\Delta_{\bf k}=0$, we have a normal metal but beyond the
arcs quasiparticles are gapped.   Open arcs also occur\cite{norm07,lee06} for the usual $d$-wave BCS spectral function (with $\mu_{\bf k}=0$ and $\Delta_{\bf k}=\Delta_0(\cos k_x - \cos k_y)$ in Eq. (\ref{specfun})) in the presence of a finite $\gamma$.  In this case, quasiparticle poles at $\omega=\pm \Delta_{\bf k}$ are smeared together to give zero-frequency peaks that trace out open arcs.

{\it The common feature of open Fermi arc models is that}
 ${\bf v}_{el}$, being perpendicular to the truncated arc,
{\it does not sweep through all in-layer directions}
 (see Fig. 2). If the arc is defined by a sudden onset of the pseudogap then, on the arc, ${\bf v}_{el}={\bf v}_b=d\xi/d{\bf k}$ the normal band velocity. In the BCS model the quasiparticle electric current is proportional to ${\bf v}_b$ everywhere.    So, in either case, the variation of ${\bf v}_{el}$ among zero-energy quasiparticles accounts for only a limited range of directions.  There are no low-energy quasiparticles that carry current in, for example, the antinodal direction.  This is why open arc FSs have similar IMR properties to quasi-1D metals.

If open arcs are short then ${\bf v}_b$ hardly changes over the arc length and all low-energy quasiparticles carry current in nodal directions.  Eq. (\ref{rhoclean}) simplifies to:
\begin{equation}
\label{uni}
\rho(B)=\rho(0)\bigg{[}1+(\Omega_C\tau)^2-\frac{(\Omega_C\tau)^4\sin^22\phi_B}{1+(\Omega_C\tau)^2}\bigg{]}.
\end{equation}
When the field is in the antinodal direction $\phi_B=0$, we have $\rho(B)\propto B^2$ at high field.  For the
nodal field orientation $\phi_B=\pi/4$ the resistance saturates at $\rho(B)/\rho(0)=2$.    These two extreme cases result from there being, respectively, none or all of the charge on the arc moving parallel to ${\bf B}$.

 In the Figures we present detailed results of representative models.  For a pocket model we use the
Energy Displaced Node (EDN)
 parametrization\cite{norm07, ng05,xia09} of Eq. (\ref{specfun}), for which $\mu_{\bf k}=\mu_0$, $\xi_{\bf k}=2t(\cos k_x + \cos k_y)$, and $\Delta_{\bf k}=\Delta_0 (\cos k_x - \cos k_y)$.  The $d$-wave BCS quasiparticle dispersion is modified by a constant shift $\mu_0$ of the chemical potential.  There is an elliptical Fermi pocket associated with the second term in Eq. (\ref{specfun}) (we include the first term in numerical calculations but it has little effect--the same goes for the Zeeman energy shifts in Eq. (\ref{rho1})).
 Into Eq. \ref{specfun} we substitute ${\bf k}\to{\bf k}-e{\bf A}$ everywhere.

In Fig. 1 we display the magnetoresistance of the EDN pocket model.  In the upper left figure the pockets are indicated relative to the normal metallic FS. (We have used $\mu_0/t=0.05$ and $\Delta_0/t=0.2$ to produce pockets of length and shape in qualitative agreement with ARPES arcs and also take $t=20k_BT=10\gamma$.  The limit $t>>k_BT>>\gamma$ is thus assumed, but results are not sensitive to parameter values within this limit, and are chosen for numerical convenience.)  The Cartesian plot shows the variation of $\rho_{zz}(B)/\rho(0)$ versus $B$ for two field orientations: along the nodal and antinodal directions.  Both show the linear behavior indicative of the 2D FS.  In the middle of the figure we have a polar plot of $\rho(B)$ for a value $\Omega_C\tau=3$.

\begin{figure}
\begin{center}
\includegraphics[width=4.8 in, height=4.2 in]{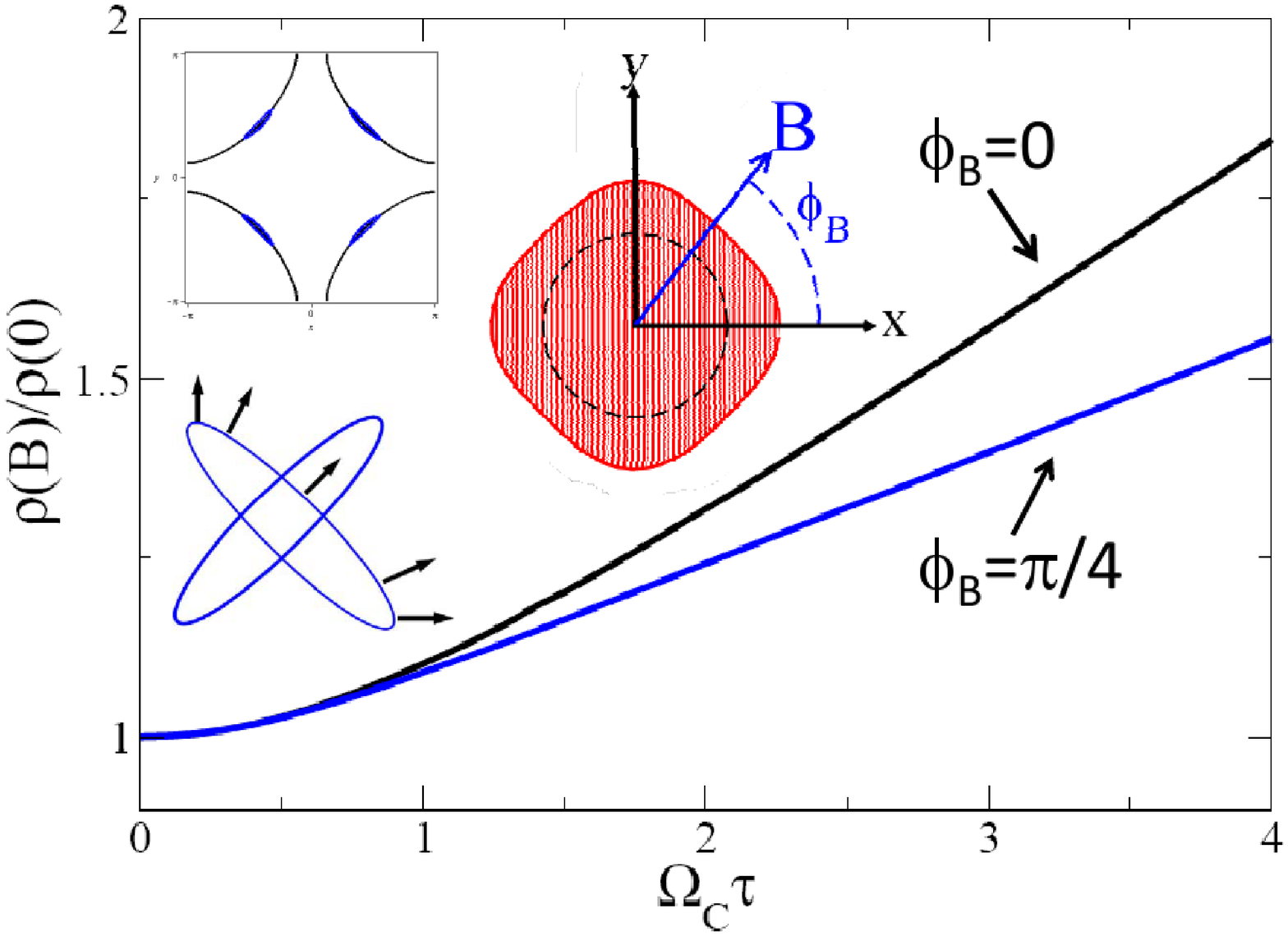}
\end{center}
\caption{Weak anisotropy of the  interlayer magnetoresistance (IMR) of Fermi pockets.
 In the upper left inset, pockets obtained from the EDN parametrization discussed in text are shown along with the normal metal FS.  In the lower left the current carried by quasiparticles on (four overlayed) pockets is indicated by arrows normal to their surface--all possible in-layer directions of current are accounted for so there are always quasiparticles carrying current parallel to an in-layer magnetic field ${\bf B}$.  In a strong field the result is a linear-$B$ dependence of IMR $\rho(B)$ for any field orientation $\phi_B$.  This is shown in the main plot, which compares the $B$-dependence (the dimensionless quantity $\Omega_C\tau$ is proportional to $B$) of $\rho(B)$ for $\phi_B=0$ and $\phi_B=\pi/4$.  A polar plot of $\rho(B)$ versus $\phi_B$ (for $\Omega_C\tau=3$, the dashed circle marking $\rho(B)=\rho(0)$) is shown in the middle.}
   \end{figure}
\begin{figure}
\begin{center}
\includegraphics[width=4.8 in, height=4.2 in]{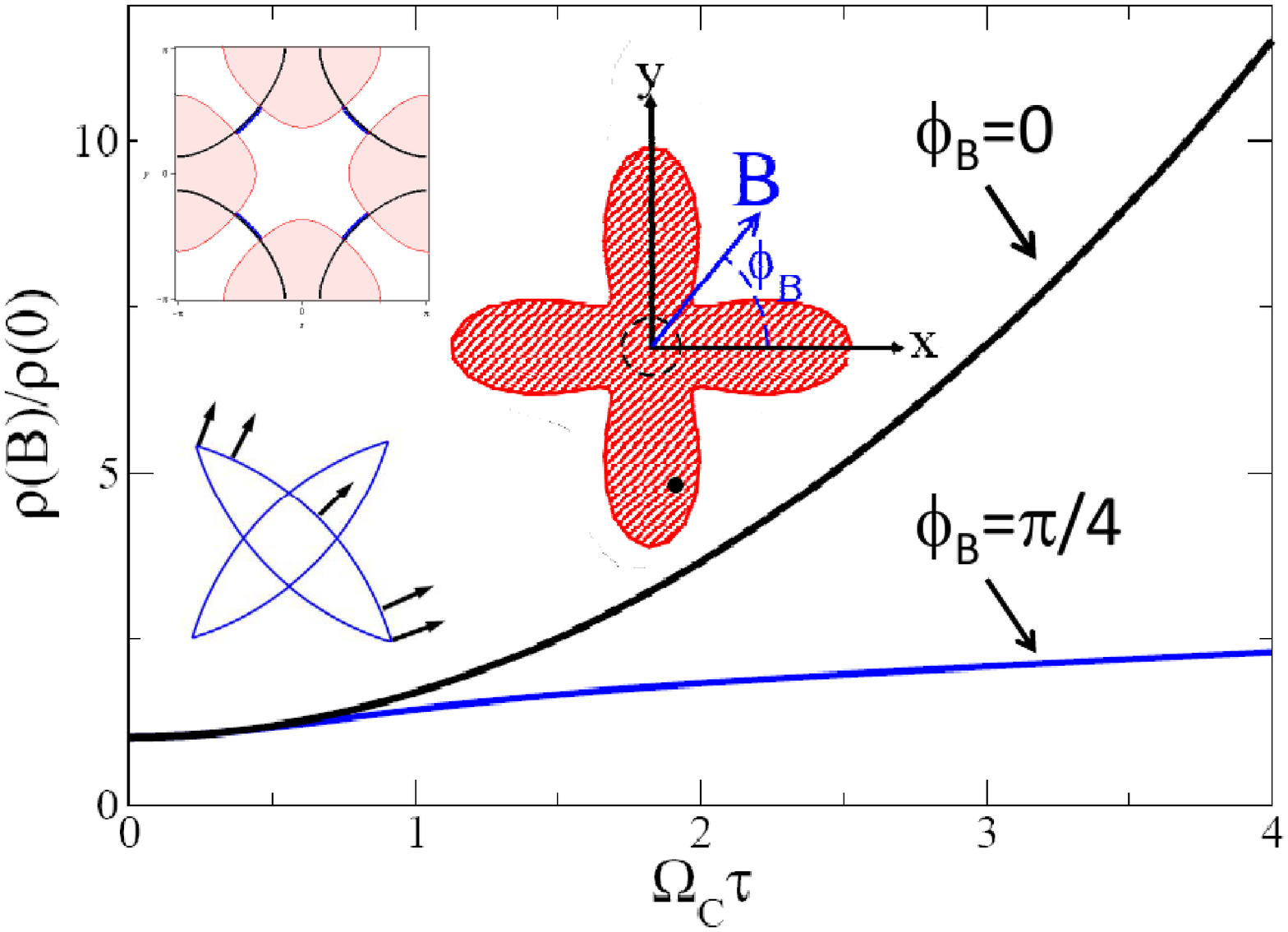}
\end{center}
\caption{Strong anisotropy of the interlayer magnetoresistance (IMR) of Fermi arcs.  If the pseudogap $\Delta_{\bf k}$ vanishes over finite-length segments of the normal-metal FS then open arcs occur.  In the upper-left figure, the shaded regions indicate where $\Delta_{\bf k}$ is present with arcs, shown as thick blue lines, in intervening regions (see text for detailed parametrization).  In the lower left, arrows represent quasiparticle current on four overlayed Fermi arcs.  Because the arcs are truncated, quasiparticles do not carry current in all possible in-layer directions (e.g. no such arrow would point in horizontal). This results in a strong dependence of the IMR $\rho(B)$ on the field orientation $\phi_B$. In main plot $\rho(B)$, plotted versus $B$, increases rapidly when $\phi_B=0$ because no quasiparticles carry current along ${\bf B}$, but it approximately saturates when $\phi_B=\pi/4$ since many quasiparticles do.  In the middle a polar plot is shown of $\rho(B)$ versus $\phi_B$ (for $\Omega_C\tau=3$, the dashed circle marks $\rho(B)=\rho(0)$).  The anisotropy is far stronger for open arcs than for the closed pockets depicted in Fig. 1.}   \end{figure}

To model arcs we multiply a $d$-wave $\Delta_{\bf k}$ by a quasi-step function of direction ( i.e. we substitute the $d$-wave $\Delta_{\bf k}$ with $\Delta_{\bf k}[\theta(\phi-\phi_0/2)+\theta(-\phi-\phi_0/2)]$ where $\phi$ is the polar angle measured from each diagonal) and use $\mu_{\bf k}=0$.  The arc length $\phi_0$ is taken to be similar to the pocket length of Fig. 1, and the magnitude of the $d$-wave gap, temperature and scattering rate remain the same.

The results are presented in Fig. 2 where $\rho(B)$ is plotted versus field strength $B$ for antinodal ($\phi_B=0$) and nodal ($\phi_B=\pi/4$)
 field orientations.  For $\phi_B=0$ the resistance increases nearly quadratically in field while for $\phi_B=\pi/4$ it shows signs of saturating (results that, though, similar to Eq. (\ref{uni}), show less anisotropy because of the finite length and curvature of arcs).  The qualitative difference between Figs. 1 and 2, both in the field dependence and anisotropy of IMR, illustrates the power of the technique for distinguishing pockets from arcs.

As mentioned above, the pure $d$-wave model produces open arcs but this model is special because the density of states depends on energy, and we need to clarify results.  While Eq. (\ref{uni}) applies when $k_BT>>\gamma$, the prefactor is strongly $T$-dependent in this clean limit (see Ref. \onlinecite{bula99}).  Also, the arcs (whose length is proportional to $\gamma$) are extremely short at temperatures $k_BT<<\Delta_0$, and not likely to produce QOs with measurable frequency.  The opposite, dirty, limit $\gamma>>k_BT$ is known to give rise to universal transport behavior\cite{lee93}.  Large values of $\gamma$ (linear in $T$ with magnitude growing to at least $\Delta_0\approx 50 $ meV) have been used to fit ARPES spectra\cite{norm07,yang06} but since only small values of $\Omega_C\tau$ could be attained if $\gamma$ was so large, the QOs cannot be attributed to BCS-type arcs in the dirty limit either.

We considered the dirty limit $\Delta_0>>\gamma>>k_BT$ (the first inequality is needed to make $\Omega_C\tau>1$ possible).  One interesting feature arises: since an entire arc is rigidly energy-shifted by the orbital effect of field (a result that follows from the fact that the field couples to ${\bf v}_b$, which varies little over the arc) {\it negative magnetoresistance can occur from purely orbital effects.}  The effect, occurring because the orbital shift of the chemical potential off the node reveals a larger density of states, is weakly dependent on field orientation.  It is less important for open arc models with a large (and $\gamma$-independent) zero-energy density of states, since the change of DOS resulting from the orbital shift is relatively small.  Since the field couples to the quasiparticle velocity on pockets, there is no corresponding energy shift.

The question of whether the QOs originate from closed pockets or open arcs can, in principle, be answered by IMR.  However, previous IMR measurements made at high magnetic field\cite{yan95,ono,krus,vede} have not revealed a strong anisotropy.  We address this discrepancy below, first noting that the observation of QOs in underdoped systems was made only in the past four years (as was the corresponding observation in overdoped systems, which reveal a large normal-metallic FS) and the improvements in sample quality that made this possible could usher in a new generation of IMR measurements as well.

IMR data of underdoped systems shows a large negative magnetoresistance, which appears to depend only weakly on field orientation \cite{krus,vede}.  Among suggestions made to explain this phenomenon, a field-dependent pseudogap ($\Delta_0$ decreases with $B$ due to Zeeman effects) has been proposed.
The primary effect of a field-dependent gap $\Delta_0$ is an isotropic drop in the magnitude of $\rho_{zz}(B)$.   A decrease in $\Delta_0$ results in pockets extending further from the zone diagonals (the pocket length being proportional to $\Delta_0$).  Since $t_\perp({\bf k})$ vanishes at nodes, lengthening the pockets increases the Fermi-surface averaged $t_\perp^2({\bf k})$ and decreases resistance.  The effect can be included by replacing $\rho_{zz}(0)$ in Eq. (\ref{rhoclean}) by a factor $\rho_{zz}^0(B)$ that depends on field strength.  For open arcs, results depend on the relationship between the magnitude of $\Delta_0$ and arc length.  However, the result for short arcs, Eq. (\ref{uni}), still holds, with any decrease in gap magnitude absorbed into the prefactor.  Thus, a field-dependent pseudogap gives negative magnetoresistance, but according to results above the relative magnitude of different field orientations would not be changed.

The interlayer matrix element favors antinodal regions, which may not be well-described by Eq. (\ref{rhoclean}) and make an additive (presumably weakly-anisotropic) current contribution.  As long as arcs are not too small, it should be possible for nodal contributions (i.e. any contribution from the spectral arc) to be extracted.  Towards this end, it may be helpful to consider thallium-cuprates, the crystal symmetry of which results in a matrix element that vanishes in antinodal directions (as well as nodal directions) \cite{huss03}.
The suppression of antinodal regions will increase the relative contribution of the spectral arcs (compared to most other cuprates where $t_\perp({\bf k})$ is maximal at antinodes).  Hence, in thallium cuprates, one need not be as concerned with the possible IMR contribution of antinodal electron pockets\cite{AFMnote}.

In conclusion, we have described calculations
of the interlayer magnetoresistance
for two qualitatively distinct classes of theoretical models for
the Fermi surface in underdoped cuprate superconductors.
These results are significant because they clearly show
that measurements of the dependence of the IMR on
the direction of the magnetic field should distinguish
between closed Fermi pockets and open Fermi arcs.

The work was supported by the Australian Research Council (DP1094395 and 
DP0710617).  We thank B.J. Powell, A.J. Schofield,
 and J. Fjaerstead for useful discussions.


\begin{thebibliography}{99}
\bibitem{AFMnote} Alternatively, if the magnetic field induces antiferromagnetic order, then (nodal hole and antinodal electrical) pockets in the zone-folded FS could provide for conventional QOs.  The resulting FS would have similar IMR properties to a model with anisotropic nodal pockets, with quantitative differences.
\bibitem{omtau} The minimum field condition for strong IMR effects is $\Omega_C\tau\geq 1$.  The condition for QOs $\omega_C\tau > 1$ is more stringent because $\Omega_C$ exceeds $\omega_C$ by a factor of the ratio $c/a$ of lattice constants (at least three).  Fields of order $50$ T have been found sufficient to achieve $\omega_C\tau>1$ in underdoped systems, so the field ranges of Figs. 1 and 2 should be experimentally possible.
\end{thebibliography}
\end{document}